\documentclass[12pt,preprint]{aastex}
\usepackage{natbib}
\usepackage{amsmath,amssymb}
\usepackage{ulem}
\usepackage{graphicx}

\begin{document}

\title{A Revised Limit of the Lorentz Factors of GRBs with Two Emitting Regions}

\author{Yuan-Chuan Zou$^1$, Yi-Zhong Fan$^2$ and Tsvi Piran$^3$}
\affil{$^1$\ School of Physics, Huazhong University of Science and Technology, Wuhan 430074, China; zouyc@hust.edu.cn}
\affil{$^2$\ Purple Mountain Observatory, Chinese Academy of Science, Nanjing 210008, China; yzfan@pmo.ac.cn}
\affil{$^3$\ Racah Institute of Physics, The Hebrew University, Jerusalem 91904, Israel; tsvi@phys.huji.ac.il}

\begin{abstract}

Fermi observations of GeV emission from GRBs have suggested that the Lorentz factor of some GRBs is around a thousand or even higher. At the same time the same Fermi observations have shown an extended GeV emission indicating that this higher energy emission might be a part of the afterglow and it does not come from the same region as the lower energy prompt emission. If this interpretation is correct then we may have to reconsider the opacity limits on the Lorentz factor  which are based on a one-zone model. We describe here a two-zone model in which the GeV photons are emitted in a larger radius than the 
MeV photons and we calculate the optical depth for pair creation of a GeV photon passing the lower energy photons shell. We find that, as expected,
the new two-zone limits on the Lorentz factor are significantly lower.  When applied to Fermi bursts the corresponding limits are lower by a factor of 
five compared to the one-zone model.  
It is possible that both the MeV and GeV regions have relatively modest Lorentz factors ($\sim 200-400$) , which is significantly softer then one zone limit
. 
\end{abstract}


\section{Introduction}

Highly relativistic motion, essential  to overcome the 
Compactness problem \citep{Ruderman75},   is a basic ingredient of  all  GRB models.  
The value of  the bulk Lorentz factor, $\Gamma$, of the relativistic outflow is of outmost interest.  
It is  essential for understanding the nature of the inner engine,  the outflow and its acceleration 
and collimation mechanisms,  the conditions at the emitting regions and the radiation mechanism. 
So far the most robust method to estimate $\Gamma$ was using the Compactness.
The   high energy photons  set an upper limit on the optical depth 
for pair production  \citep{feh93,Piran95,wl95,
ls01}.  The observations of GeV photons from several bursts enabled 
the Fermi team to set very high ($\Gamma \ge 1000)$
lower limits on $\Gamma$ for those GRBs \citep{Abdo09a,Abdo09b,Ackermann10}. 
While other methods to estimate or limit $\Gamma$ depend on various assumptions  (see e.g. \citet{zp10}), the compactness limit seems to be independent of any model assumptions and hence it is considered to be the most robust one. 

However, 
recent  Fermi observations have also shown that the onset of higher energy ($\sim 100$ MeV to a few  GeV - denoted hereafter GeV)
emission lags after the onset of the lower energy ($\sim 100$ keV to a few MeV - denoted hereafter MeV) prompt 
emission \citep{Abdo09a,Abdo09b,Ackermann10,ggnc10}. Fermi confirmed earlier EGRET results that 
the GeV  emission also  lasts longer than the MeV emission  \citep{hd94,fm95,gd03}. 
These two facts suggest the possibility of a different origin for the 
MeV and the GeV emission. This would be the case, for example, if 
the GeV emission arises from an external shock afterglow (Kumar \& Barniol Duran 2009, 2010; Ghisellini et al. 2010; Piran \& Nakar 2010;  Gao et al. 2009), from multi-zone internal shocks \citep{xfw08,Aoi09,zlb10} or if the MeV emission is 
the quasi-thermal radiation of the baryonic outflow while the GeV emission is  
mainly from the subsequent  internal shocks\footnote{The synchrotron radiation of electrons accelerated by such internal shocks may peak at eV energies while the inverse Compton radiation can give rise to a significant GeV component. In such a scenario, the
GeV emission and energetic ultraviolet/optical flare are tightly correlated.}.

\begin{figure}
\includegraphics[width=\textwidth]{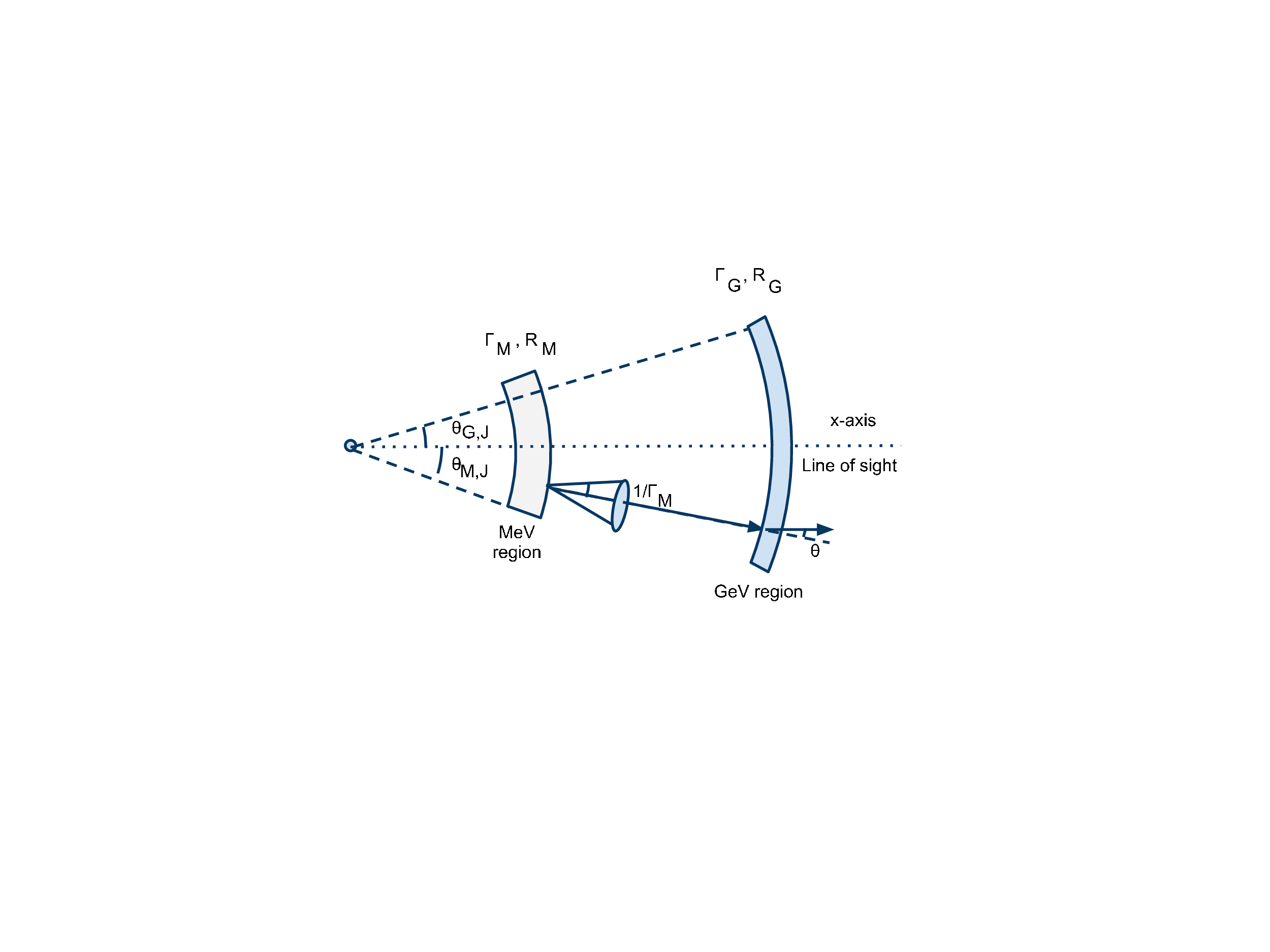}
\caption{A schematic diagram of  the two-region scenario. GeV photons are produced at a larger radius. The MeV shell engulf this radius when the GeV photons are produced. At $R_M$ the MeV photons beamed into an angle $1/\Gamma_M$. The GeV photons can reach us only if they are directed in the line of sight, which is along the x-axis.}
\label{Fig.geometry.1}
\end{figure}

The compactness limits on the Lorentz factor  are based, however, on an implicit  assumption that the MeV and the GeV photon arise from the same region.  
We show here that the relaxation of this assumption  reduces significantly the estimated lower  limit on $\Gamma$
 \citep[see also][]{Aoi09,Li10,zlb10}.
The existence of two regimes leads to a rich variety of possibilities. We consider  in the following the most natural configuration,  which is also consistent with the temporal delay of the GeV emission, the MeV emission is produced at lower radii (say via internal shocks) and the GeV emission is 
produced at a larger radius (say via external shocks) - see Fig. \ref{Fig.geometry.1}. 
Other geometrical options that we don't consider are that the GeV emission is emitted at a lower radius than the MeV emission or that  the MeV and the GeV emission are produced at  different angular regions.  Our estimates don't depend on the origin of the emission (e.g. internal or external shock) but just on the overall geometry of the system.  
We calculate the optical depth of a GeV photon passing through the MeV photons shell and  we obtain new compactness limits on $\Gamma$.


\section{The Model}
\label{Model}

\begin{figure}
\includegraphics[scale=1]{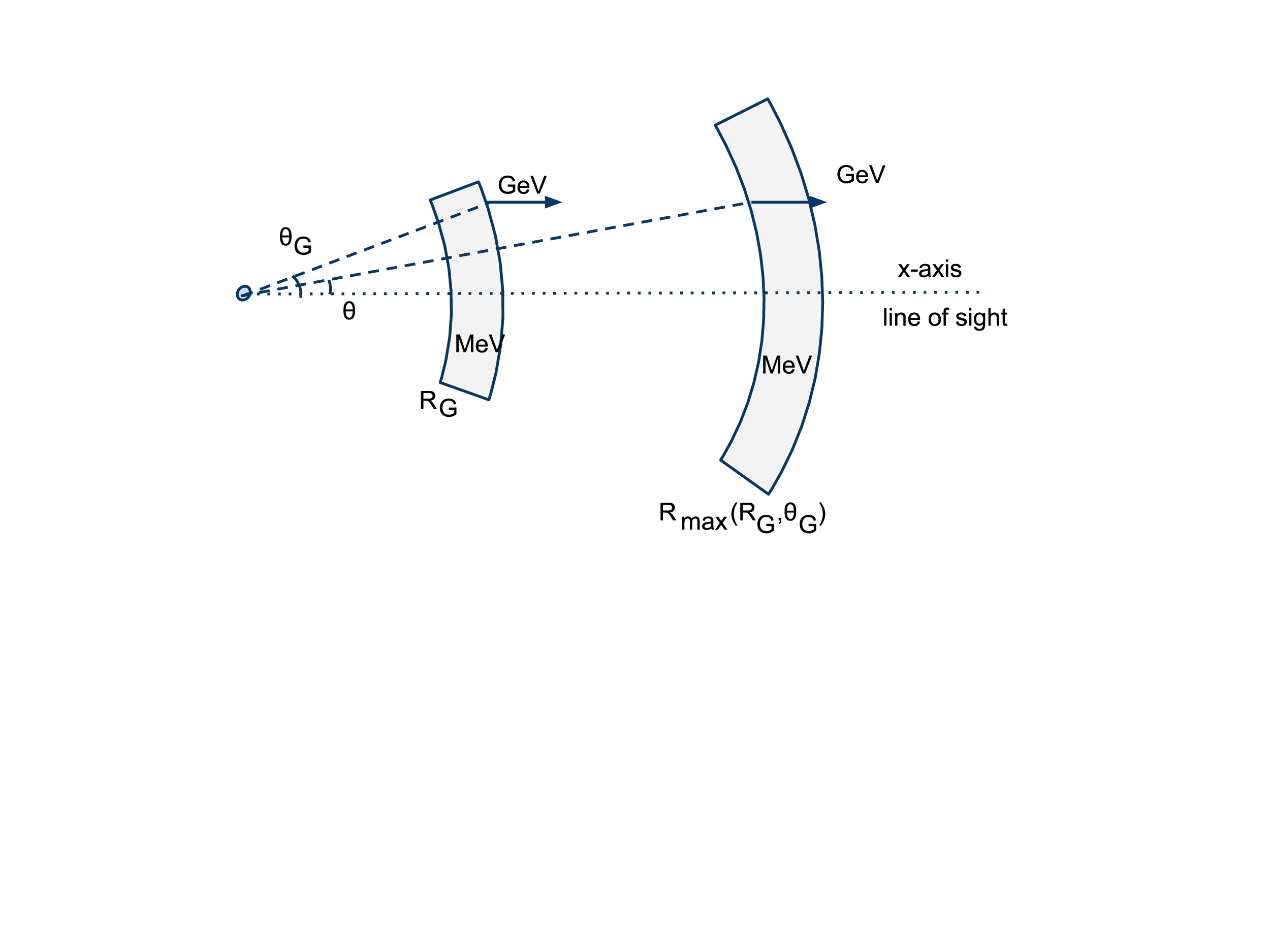}
\caption{A schematic diagram  of the geometry of the GeV photon trajectory within the MeV shell. The GeV photons meets the MeV photon shell (with a width $\Delta_{M}\simeq  c \delta t_M$) at $R_G$ and it leaves the shell at $R_{\rm G,\max}$.}
\label{Fig.geometry.2}
\end{figure}

Consider two emitting regions denoted by $M$, for MeV and $G$, for GeV  respectively (see Fig. \ref{Fig.geometry.1}). The MeV (GeV) emission region  has a radius $R_M$ ($R_G$) (for simplicity we consider emission from thin shells) with   $R_M \ll R_{G}$,  a Lorentz factor $\Gamma_M$ ($\Gamma_G$) and an angular width  $\theta_{M,J}$ ($\theta_{G,J}$).  We assume that the MeV and GeV jets are aligned and both are pointing toward us.  In this configuration a GeV photon passes through a shell of MeV photons on its way to the observer (see Fig. \ref{Fig.geometry.2}). 
Our goal is  to estimate the optical depth for pair production of a GeV and  an MeV photons shell. 
The width of the MeV photons shell is\footnote{\cite{zlb10} consider erroneously an MeV shell width of $R/2\Gamma$ (see their Eq. (10)).}: 
\begin{equation}
\Delta_M  \simeq c T_{90} /(1+z),
\label{Mshell}
\end{equation}
 where $T_{90}$ is the observed duration of the MeV pulse and $z$ is the redshift of the burst. 
A GeV photon,  emitted  along the $x$ axis at 
$ R_G $  and $\theta_G$ is immersed in MeV photons until it  leaves the MeV shell at $R_{\max}(R_G,\theta_G)$ (see Fig. \ref{Fig.geometry.2}):
\begin{equation}
R_{\max}(R_G,\theta_G) =  \frac{R_G}{1 -{2 c T_{90}}/[{(1+z)\theta_G^2R_G}]},
\end{equation}
for $\theta_G^2 > {2 c T_{90}}/[{(1+z)R_G}]$, while $R_{\max}(R_G,\theta_G) \rightarrow \infty$ for $\theta_G^2 \leq {2 c T_{90}}/[{(1+z)R_G}]$. 

Assume, for simplicity, that the emitted flux of the MeV photons is constant  over time then the number density of the MeV photons is:
\begin{equation}
n_{0,{MeV}} (R) \simeq \frac{L_M}{(1+z)^2 E^2_{P}  4 \pi R^2  c},
\end{equation}
where $L_M$ is the isotropic equivalent MeV luminosity,
$E_{P}$ is the observed peak energy and $R_G< R< R_{\max}(R_G,\theta_G)$ is the radius.
The spectrum of the MeV photons is described by the Band function:
\begin{eqnarray}
 n_{MeV}(E_M) \simeq n_{0,{MeV}} (R)  \left\{  
  \begin{array}{ll}
   \left(\frac{E_M}{E_P}\right)^{-\alpha_{_M}} \phantom{00} & E<E_P , \\
   \left(\frac{E_M}{E_P}\right)^{-\beta_M} \phantom{00} & E_p<  E < E_{M,{\max}},
  \end{array}
   \right. 
\end{eqnarray} 
where, in view of the two region model,  the MeV component has an upper limit to the energy: $E_{M,{\max}}$. 
Fortunately  $\tau(E_G)$ is insensitive to the exact  value of $E_{M,{\max}}$.

At $R_M$ The MeV photons are beamed with an angular width $1/\Gamma_M$ along the radial direction outwards. 
Since $R_G> R_M$ the MeV photons travel almost radially outwards.
Relative to  the GeV photons, that move along the $x$ axis, the angular width of the MeV photons is of order $R_M/(R \Gamma_M)$ and 
it decreases with $R$. This leads to two  angular regimes: Along the $x$ axis   (for $\theta_G <R_M/(R \Gamma_M)$), the MeV photons have  very small angles $\le R_M/(R \Gamma_M)$ relative to the GeV photons. This  leads to a very small optical depth along the axis. For $\theta_G >R_M/(R \Gamma_M)$ the angular spread of the MeV photons can be neglected and the typical angle between the  MeV and the GeV photons is simply $R_G \theta_G/R$. 

More generally, the angle between an  MeV photon emitted  radially outwards at $(R_M, \theta_M)$ 
and a GeV photon emitted parallel to the $x$ axis at $(R_G, \theta_G)$  is:
\begin{equation}
\theta = \arctan \frac{\sqrt{(R_M \sin \theta_M)^2 + (R_G \sin \theta_G)^2 - 2 (R_M \sin \theta_M)(R_G \sin \theta_G) \cos \phi}}{R_G \cos \theta_G - R_M \cos \theta_M + \Delta	}
\end{equation}
 where $\Delta=R-R_G$ is the distance the GeV photon travels before it collides with the MeV photon and $\phi$ is the angle of a given MeV photon relative to the axis parallel to the GeV photon axis.

A GeV photon, with an (observed) energy $E_G$,  and an MeV photon, with { an  (observed) energy $E_M$,} can produce a pair if  $E_M \ge E_{M,\min}$, where:  
\begin{equation}
E_{M,\min}  = \frac{ 2 (m_e c^2)^2 }{(1+z)^2 E_{G} (1-\cos \theta)} \simeq \frac{4(m_e c^2)^2 }{(1+z)^2 \theta^2 E_{G}} . 
\label{eq:e_min}
\end{equation}
{ The cross section} behaves like \citep[e.g.][]{JR80}:
\begin{equation}
 \sigma = \frac{3}{16} \sigma_T (1-\hat \beta ^2) \left[(3-\hat \beta^4)\ln \frac{1+\hat \beta}{1-\hat\beta}-2\hat\beta (2-\hat\beta^2)\right],
 \label{eq:sigma}
\end{equation}
where 
\begin{equation}
\hat \beta = \sqrt{1-2\frac{m_e c^2}{(1+z)E_M} \frac{m_e c^2}{(1+z)E_G} (1-\cos\theta)^{-1}},
\label{eq:beta}
\end{equation}
{ $\hat \gamma = 1/\sqrt{1-\hat \beta^{2}}$ and
$\sigma_T \simeq 6.65\times 10^{-25} {\rm cm^2}$ is the Thompson cross section.}

We can estimate now the overall optical depth for the  GeV photon:
\begin{equation}
\tau(\theta_G,E_G) = \int_{R_G}^{R_{\max}(R_G,\theta_G)} dR  \int_{0}^{\theta_{M,j}} d\theta_M \int_0^{2\pi} d\phi \int_{E_{M,\min}}^{E_{M,\max}} dE_M  \frac{d ^3 n_{MeV}}{dE_M d\theta_M d\phi} \sigma (1-\cos \theta)
\label{eq:tau_theta}
\end{equation}
{The effective optical depth for an observed photon is  averaged over all angles:
\begin{equation}
\exp [-\bar{\tau} (E_G)] = \frac{\int_{0}^{\theta_{G,j}}  \mathcal{D}^{(3+\beta_G)} e^{-\tau(\theta_G,E_G)}     \theta_G d\theta_G}{ \int_{0}^{\theta_{G,j}} \mathcal{D} ^{(3+\beta_G)}    \theta_G d\theta_G},
\label{eq:e_tau_complete}
\end{equation}
where $\mathcal{D}=1/(\Gamma_G({1-\beta_{\rm bulk,G} \cos \theta_G}))$ is the Doppler factor, $\beta_{\rm bulk,G}=\sqrt{1-1/\Gamma_G^2}$ is the bulk velocity and $\beta_G$ is the photon index of the GeV emission.}
It is interesting to note that Eq. (\ref{eq:tau_theta}) is independent of $\Gamma_G$. The overall dependence of $\tau$ on $\Gamma_G$ arises from Eq. 
(\ref{eq:e_tau_complete}) where $\Gamma_G$ determines (via $\mathcal{D}$) the effective width of the integration.

\section{A Simplified model}\label{sec:simplified}

We can simplify  the above model and obtain an almost analytic formula by making a few approximations.  We show later that the full numerical results
indeed agree with these formulae. 
First, we approximate the MeV spectrum  in the relevant energy range using a single power law: $N_{MeV} (E_M) \propto E_M^{-\alpha_{_M}}$. Second, we approximate the {\
cross section (Eq. (\ref{eq:sigma}))  as  $\sigma\simeq  \sigma_T/(3\hat \gamma^2)$ (This form allows us
for an analytic integration. The numerical factor is chosen by
comparison of the analytic approximate results with the full numerical
solution.).}
Third, we divide the analysis to two regimes: For $\theta_G >  R_M/(R \Gamma_M)$ all the MeV photons move radially and the collision angle is simply $\theta\simeq \theta_M$.
For $\theta_G < R_M/(R \Gamma_M)$ the scatter in the directions of the  MeV photons is important and we approximate { the collision angle} as $\theta\approx R_M/(R \Gamma_M) $.

Define $x\equiv {R}/{R_G}$ and  $x_{\max} \equiv {R_{\max}}/{R_G} =1 /[{1-2 \Delta_{M} / (\theta_G^2 R_G})]$. For large angles the  collision angle between the MeV and GeV photons is $\theta \simeq {R_G \theta_G }/{R} = {\theta_G}/{x}$.
The minimal energy of the MeV photon for pair production  is: 
\begin{equation}
E_{M,\min} \simeq 
\frac{4(m_e c^2)^2}{(1+z)^2E_G \theta_G^2} x^2.
\end{equation}
The number density of the MeV photons:
\begin{equation}
N(E_M) \simeq n_{0,MeV}(R) (\frac{E_M}{E_P})^{-\alpha_{_M}} = \frac{(1+z)^{2\alpha_{_M}-2}L_M E_G^{\alpha_{_M}} E_P^{\alpha_{_M} - 2} \theta_G^{2\alpha_{_M}}}{4\pi R_G^2 c (2\pi m_e c^2)^{2\alpha_{_M}}} x^{-(2\alpha_{_M}+2)} y^{-\alpha_{_M}},
\end{equation}
 where $y\equiv {E_M}/{E_{M,\min}} = \hat{\gamma}$.
{ Collecting the above expressions:
\begin{eqnarray}
 \tau(\theta_G) &=& \int_{R_G}^{R_{\max}(R_G,\theta_G)}{\rm d} R \int_{E_{M,\min}(\theta,R)}^{E_{M,\max}} {\rm d}E_M \cdot N(E_M) (1-\cos\theta)	 \sigma_T/(3{\hat \gamma}^2) 
\nonumber \\
&\simeq & \frac{(1+z)^{2\alpha_{_M}-3}L_M \epsilon_G^{\alpha_{_M}-1} \epsilon_P^{\alpha_{_M}-2} \sigma_T \theta_G^{2\alpha_{_M}}}{6\cdot 
 4^{\alpha_{_M}} \pi  R_G m_e c^3 \alpha_{_M}} \nonumber \\
&& \left[\frac{1}{2\alpha_{_M}+1}\left(1-x_{\max}^{-(2\alpha_{_M}+1)}\right) - \left(\frac{1}{4} (1+z)^2 \epsilon_{\max}\epsilon_G \theta_G^2\right)^{-\alpha_{_M}} (1-\frac{1}{x_{\max}}) \right],
 \label{eq:tau_theta_G_sim}
\end{eqnarray}
where $\epsilon_G \equiv {E_G}/{m_e c^2}$  and $\epsilon_{\max} \equiv {E_{M,\max}}/{m_e c^2}$. }

 We use $R_G=\eta 2\Gamma_G^2 c T_G/(1+z)$ to express $R_G$ in terms of $T_G$ and $\Gamma_G$, where  $T_G$ is the time when the GeV photon is observed. Generally $\eta \le 1$ with $\eta = 1$ for an external shock  and $\eta<1$($\sim 0.01$) for an internal shock. As we see later the result is not very sensitive to $\eta$. 
Realizing that the GeV photons are mainly coming from $\theta_G \sim 1/\Gamma_G$, we can 
invert now this equation to obtain a rough estimate of the  minimal Lorentz factor. The solution will be consistent if $\theta_G < R_M/(R_G \Gamma_M)$ namely if $\Gamma_G > R_G \Gamma_M /R_M$. To do so we assume that the factor in the square brackets of Eq. (\ref{eq:tau_theta_G_sim}) is  order of unity. { We obtain:
\begin{eqnarray}
\Gamma_{G,\min} &\approx & \left[\frac{(1+z)^{2\alpha_{_M}-3}L_M \epsilon_G^{\alpha_{_M}-1} \epsilon_P^{\alpha_{_M}-2} \sigma_T }{12\cdot 4^{\alpha_{_M}} \pi  m_e c^4 \eta T_G\alpha_{_M}}\right]^{\frac{1}{2\alpha_{_M}+2}}  \nonumber \\
& \simeq & 34 \left( \frac{30 \times 10^{4\alpha_{_M}}}{\alpha_{_M} 34^{2\alpha_{_M}+2}}\right)^{\frac{1}{2\alpha_{_M}+2}}
\left[ \left(\frac{1+z}{2}\right)^{2\alpha_{_M}-3} L_{M,51} \epsilon_{G,4}^{\alpha_{_M}-1} \epsilon_{P,0}^{\alpha_{_M}-2}  \eta^{-1} T_{G,2}^{-1}
\right]^{\frac{1}{2\alpha_{_M}+2}},
\label{eq:Gamma_G_min1}
\end{eqnarray}
where the notation $Q_x=Q/10^x$ is used  and $\epsilon_{P} \equiv E_P/m_e c^2$. For $\alpha_{_M}=2$, the overall coefficient  is 34. } Note the weak 
(with a power ${-1/(2\alpha_{_M}+2)} \sim -1/6$ to $-1/2$) dependence of $\Gamma_{G,\min}$ on $\eta$.

For   $\theta_G \le R_M/R \Gamma_M $ the typical collision angle of the MeV photons is $\theta \lesssim R_M/(R \Gamma_M)$. 
{ The  optical depth is similar to Eq. (\ref{eq:tau_theta_G_sim}), with  $\theta\sim R_M/(R \Gamma_M)$ replacing  $\theta_G$ 
and $x_{\max} \rightarrow \infty$:
\begin{eqnarray}
\tau(\theta_G \le R_M/R \Gamma_M) &\approx& \tau(\theta_G=0) \simeq 
 \frac{(1+z)^{2\alpha_{_M}-3}L_M \epsilon_G^{\alpha_{_M}-1} \epsilon_P^{\alpha_{_M}-2} \sigma_T R_M^{2\alpha_{_M}}}{6 \cdot  
4^{{\alpha_{_M}}} \pi  R_G^{2\alpha_{_M}+1} m_e c^3 \alpha_{_M} \Gamma_M^{2\alpha_{_M}}} 
\nonumber \\
&& \times \left[\frac{1}{2\alpha_{_M}+1}- \left(\frac{1}{4} (1+z)^{2} \epsilon_{\max}\epsilon_G (\frac{R_M}{R_G \Gamma_M})^{2}\right)^{-\alpha_{_M}} \right].
 \label{eq:tau_theta_M}
\end{eqnarray}
Interestingly,} this formula depends on $\Gamma_M$ and it is independent of $\Gamma_G$.  
For typical values, $z=1$, $\alpha_{_M}=2$, $E_{\max}=E_G=10$GeV, $E_P=1$MeV, $L_M= 10^{51}$erg, $R_M=10^{15}$cm, $R_G= 10^{16}$cm, {we obtain $\Gamma_{M,\min}\simeq 8$,} which provides only a very week constraint on the  bulk Lorentz factor of the  MeV region. 
If we set $R_G=R_M$, this expression reduces to the one zone case, where the relevant collisions occur at $\theta=1/\Gamma_M$ in the observer's frame. 
Taking $R_M=R_G=2\Gamma_M^2 c \delta T$, 
where $\delta T=0.1$s is the typical duration of $\gamma$-ray pulse,    { $\Gamma_M \ge  200$,  consistent  with \cite{ls01}.}

\section{Results}
\label{Results}

\begin{figure}
\includegraphics[width=\textwidth]{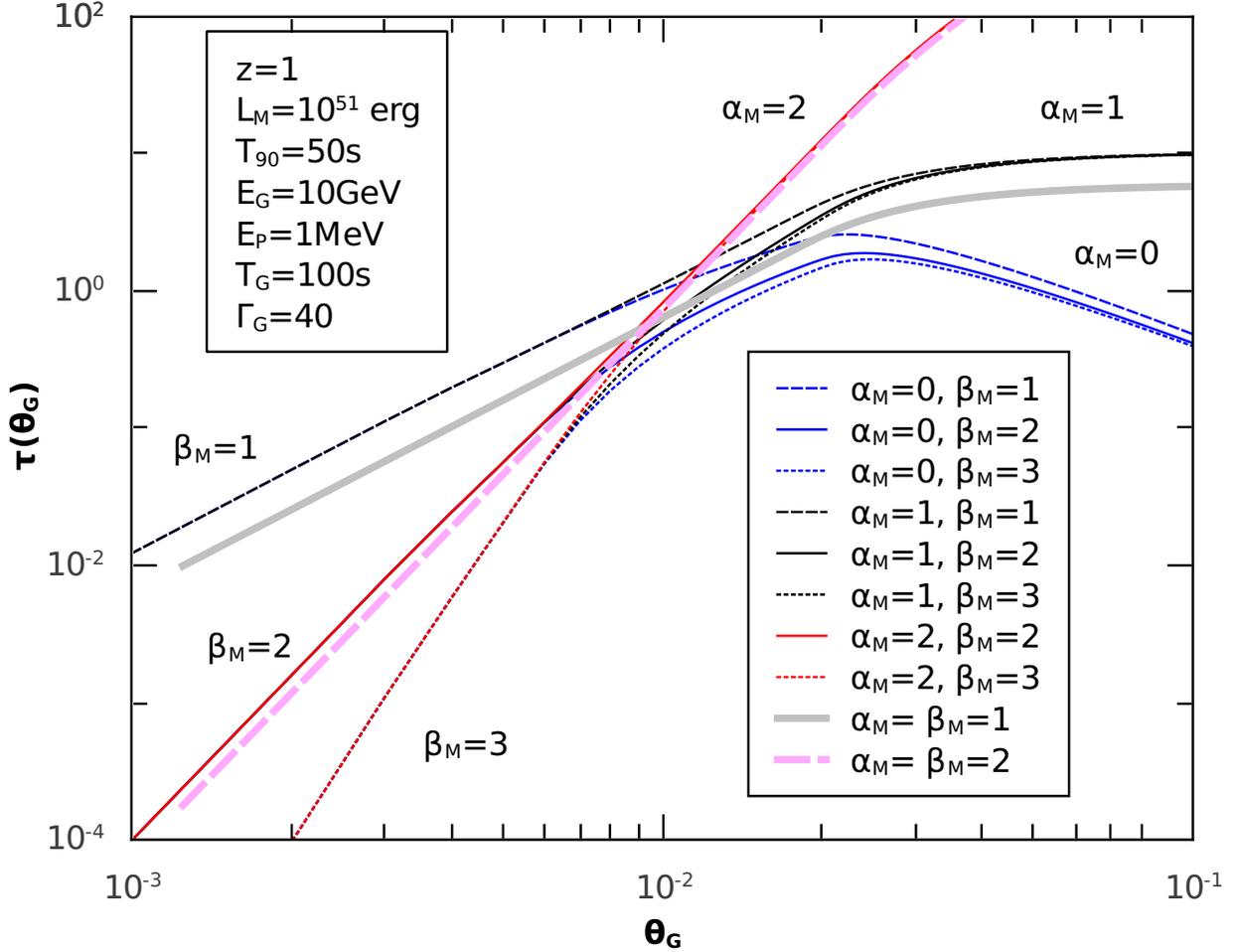}
\caption{The optical depth for a  photon  with $E_G=10$GeV as a function of $\theta_G$. Different lines correspond to different spectral indices. The parameters of the lines are encoded using a combination of colors and line types. $\alpha_{_M}=0,1,2$ are represented as blue, black and red respectively. $\beta_M=1,2,3$ are represented as dashed, solid and dotted lines respectively. The two thick lines are the analytical solution from Eq. (\ref{eq:tau_theta_G_sim}) for $\alpha_{_M}=\beta_M=1,2$.  
}
\label{Fig.tau-theta}
\end{figure}

Fig. \ref{Fig.tau-theta} depicts the dependence of $\tau$ on $\theta_G$ for a set of  typical parameters: $z=1$,  $\delta t_M=50$s and $\Delta_M=7.5\times 10^{11}$cm, $E_G=E_{\max}=10$GeV, $E_P=1$MeV, $L_M= 10^{51}$erg, and $T_G=100$s. $\beta_G$ is always chosen to be 2. 
$\bar {\tau} \simeq 1$ for these parameters (with $\alpha_{_M}=1$, $\beta_M=2$)  for $\Gamma_{G,\min} = 40$.
Note that numerical experiments reveal that to obtain the average effective optical depth using Eq. (\ref{eq:e_tau_complete}) we have to integrate up to $\sim 4 /\Gamma_G$. 
The results in this figure depict both the  full 
(Eq. (\ref{eq:tau_theta}))  and the simplified (Eq. (\ref{eq:tau_theta_G_sim})) calculations.
The later are depicted by the two thicker lines which are almost superposed on  the corresponding numerical results, showing consistency of   the analytical solution  with the full numerical one. 
As the angle increases, the small angle approximation ($\sin \theta \simeq \theta$) breaks down and the approximate solution slightly diverges from the numerical one.  
However, the deviation is small and it  usually takes place in a regime that is not critical for the overall optical depth.

One can clearly see different segments of  power law  dependence of the optical depth on 
$\theta_G$. For a given GeV photon $\theta_G$ determines the  colliding angle and hence 
$E_{M,\min}$. As most  photons are in the lowest energy and the  cross section is largest near 
$E_{M,\min}$, the optical depth is dominated by the low energy photons at  $E_{M,\min}$ and 
the photon index can be simply taken the one at  $E_{M,\min}$. 
For small angle, $E_{M,\min}>E_p$  (see  Eq. (\ref{eq:e_min})) and  the effective spectral slope is $\beta_M$. 
For large angles $E_{M,\min}<E_p$ and   the effective spectral slope is $\alpha_{_M}$. 
Correspondingly, lines with the same  $\beta_M$ coincide at small angles while lines with the 
same $\alpha_{_M}$ conicide at large angles. These results suggest that the single power law approximation for the spectrum is useful. 
$E_{M,\min}$ determines the relevant spectral, $\alpha_{_M}$ or $\beta_M$.

A second transition takes place  when $x_{\max}$ approaches unity, namely the width of the interaction region is small compared to $R_G$.
According to Eq. (\ref{eq:tau_theta_G_sim}), $\tau(\theta_G) \propto \theta_G^{2\alpha_{_M}}$ for $x_{\max} >> 1$ and $\tau(\theta_G) \propto \theta_G^{2\alpha_{_M}-2}$ for $x_{\max} \sim 1+ 2\Delta_M/(\theta_G^2 R_G) \rightarrow 1$. The transition between the two takes place as 
$\theta_G \approx  \sqrt{2\Delta_M/R_G} \sim \sqrt{2 c T_{90}/(2\Gamma_G^2 c T_G)} \sim \sqrt{T_{90}/(\Gamma_G^2 T_G)}$.  For some parameters the two transitions may coincide to one and we have chosen $L_M$ so that the two transitions are clearly seen. 

\begin{figure}
\includegraphics[scale=1]{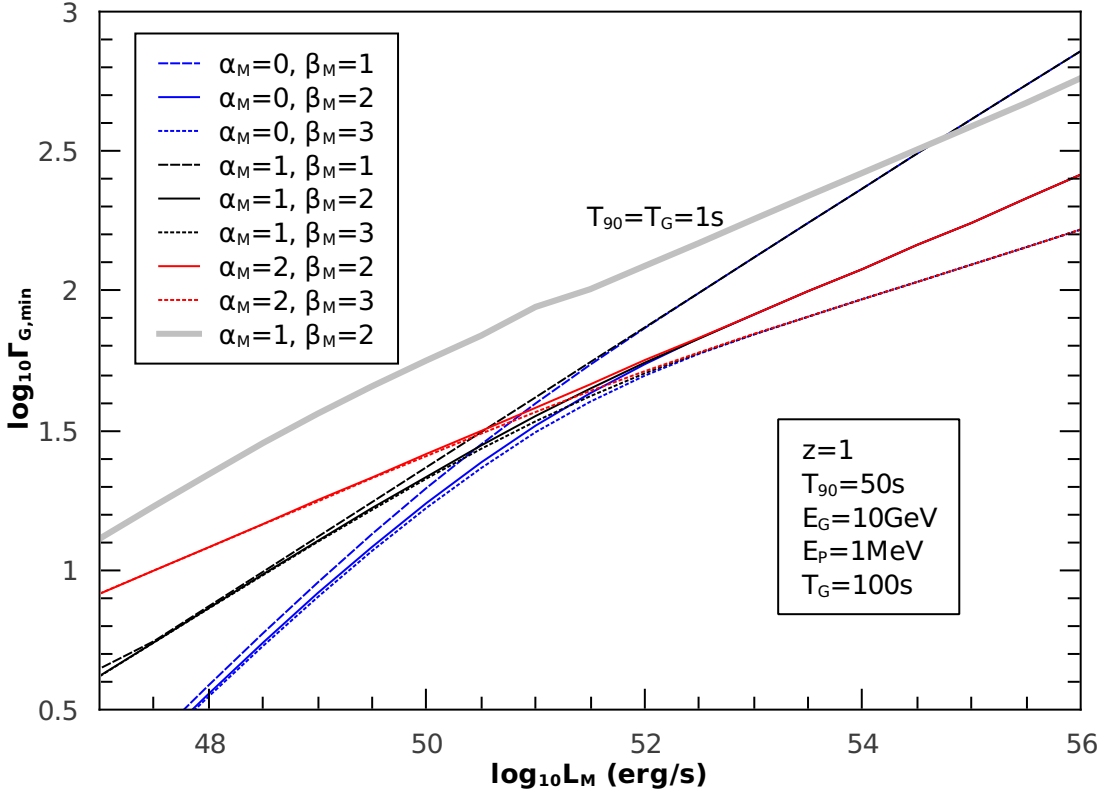}
\caption{The minimal Lorentz factor $\Gamma_{G,\min}$ as a function of the MeV luminosity, $L_M$. Different curves correspond to  different photon indices. The parameters of the lines are encoded using a combination of colors and line types. $\alpha_{_M}=0,1,2$ are represented as blue, black and red respectively. $\beta_M=1,2,3$ are represented as dashed, solid and dotted lines respectively. 
The thick grey line has  $T_{90}=T_G=1$s, corresponding to a short GRB.
}
\label{Fig.Gamma_G-L}
\end{figure}

Fig. \ref{Fig.Gamma_G-L} depicts  $\Gamma_{G,\min}$ as a function of the MeV luminosity $L_M$. As one can expect (see Eq. (\ref{eq:Gamma_G_min1})) $\Gamma_{G,\min}$ increases with $L_M$. The optical depth for  $T_G=1$s (thin solid line, $\eta=1$) is much larger than the other ones for which $T_{G}=100$s, again in agreement with  Eq.(\ref{eq:Gamma_G_min1}).
If $\alpha_{_M}=\beta_M$, $\Gamma_{G,\min}$ increases with $L_M$ as a single power law. This is consistent with  Eq.(\ref{eq:Gamma_G_min1}). For  $\alpha_{_M} \neq \beta_M$, the relationship breaks into two  power law segments that are dominated by different parts of Band spectrum of the MeV photons. As the luminosity increases the Lorentz factor increases, and the dominant contribution to the optical depth arises from $\theta_G \approx 1/\Gamma_G$, with a lower effective $\theta_G$,  $E_{M,\min}$ increases and hence for large values of $L_M$ the opacity is dominated by the high energy spectral slope, $\beta_M$. Conversely,  for low values of $L_M$ the opacity is dominated by $\alpha_{_M}$. The transition { takes place at $E_{M,\min} \approx E_p$. }

\begin{table} \footnotesize
\caption{Limits on Fermi LAT busts}
\begin{tabular}{lccccccccccc}
\hline 
GRB & z & $T_{90}$(s) & $\alpha_{_M}$ & $\beta_M$ & $E_P$(MeV) & $L_{\rm iso}$ (erg/s) & $E_G$(GeV) & $T_G$ & $\Gamma_{G,\min}\ ^{*}$ & $\Gamma_{\min} $ & ref
\\ \hline 
080916C & 4.35 & 66 &1.02 &2.21 &1.17 & $ 7 \times 10^{53}$ & $ 13.22$ & 40 & {\bf 193}(414) & 880 & 1
\\
090510 & 0.903 & 0.5 & 0.48 & 3.09 & 5.1 & $4.6\times 10^{53}$ & 3.4 & 0.5 & {\bf 150} (277) & 1200 &  2
\\
090902B & 1.822 & 30 & 0.61 & 3.87 & 0.8 & $3.2\times 10^{53}$ & 33.4 & 82 & {\bf 120} (218)& 1000 & 3,4 
 \\
090926A & 2.1062 & 20 & 0.693 & 2.34 & 0.27 & $4.2\times 10^{53}$ & 19.6 & 26 & {\bf 150} (318)& 1200 &  5 
\\ \hline 
\multicolumn{12}{l} {Shown are the low energy spectral parameters $\alpha_M$, $\beta_M$ and $E_P$ as well as the luminosity, $L_{iso}$ and the energy, $E_G$
}
\\
\multicolumn{12}{l} {and time $T_G$  of the highest  energy  GeV photon as well as the two zone, $\Gamma_{G,min}$, and the single zone, $\Gamma_{min}$ ,limits.
}
\\
\multicolumn{12}{l}{(1) \citet{Abdo09a}; (2) \citet{Ackermann10}; (3) \citet{Abdo09b}; (4) \citet{dePalma09};
}
\\
\multicolumn{12}{l}{(5) \citet{Swenson10}
}
\\
\multicolumn{12}{l}{ $^{*}$ in bold face is the value for $\eta=1$ and brackets is the value for $\eta=0.01$ 
}
\\ \hline
\end{tabular}
\label{tab:Gamma}
\end{table}

The two zone limits, $\Gamma_{G,\min}$, for  four  Fermi bursts are shown in table \ref{tab:Gamma} together with the single zone limits, $\Gamma_{\min}$, and some parameters of the bursts. 
 While the single zone limits, $\Gamma_{\min}$, are of order 1000 and even larger, the two zone limits are around 200 (400), for $\eta=1~(0.01)$. 
It should be stressed that in 
two of these bursts, GRB 080916c and  GRB 090510, the highest energy GeV photon used to determine the single zone limit is  
coincident  with a large  MeV flux,  while an 11 GeV photon is  contemporaneous in GRB 090902b with an MeV spike.  Still in both 
GRB 090902b and GRB 090926a GeV photons are observed after the end of the prompt MeV ($T_G> T_{90}$).  Thus, it is not clear whether the single zone of the two zone limit should be used.

If the GeV photons are from an external shock, another  direct constraint on $\Gamma_G$  arises from the dynamics of external shock: $\Gamma_G \simeq 240 E_{k,52}^{1/8} n_0^{-1/8} (1+z)^{3/8}T_{G,P,1}^{-3/8}$ \citep{sp99}. With observed GeV peak  emission time $T_{G,P}$ of $\sim 10$s  \citep{ggnc10},  $E_k\sim E_{\gamma,\rm iso} \sim 10^{55}$erg and a  circum-burst density $n\sim1 {\rm cm^{-3}}$, we obtain $\Gamma_G \sim 740$. This value is larger than the lower limits obtained from the two-zone compactness estimate. However, it is quite uncertain,  in view of the uncertainty in the
determination of $T_{G,P}$.

\section{Conclusions}
\label{conclusions}

Following various indications that the high energy (GeV) emission in GRBs is produced in a different region than the lower energy (MeV) emission we  derived  here  revised compactness limits on the Lorentz factor of  GRB outflows within a two-zone model. We considered the ``natural" model in which the GeV emission is produced at larger radii than the MeV emission. This would arise, for example, if the MeV emission is produced by internal shocks and the GeV emission by the afterglow, as has been suggested recently by several authors \citep{kb09,kb10,ggnc10}. We calculated the 
optical depth for pair production by a GeV photon passing through the MeV photons shell. 
Our results reduce to the one-zone model when the emission region of the MeV and GeV coincide. 

Collisions between the GeV and MeV photons occur in the two-zone model at larger radii than the prompt emission radius,  the density of the MeV emission is  smaller (than in the prompt emission regime) and the MeV photons are more collimated along the line of sight. Consequently, the optical depth is smaller compared to the one-zone case and the compactness constraint on the Lorentz factor becomes weaker. The new constraint that we find is only for the Lorentz factor of the GeV region. 
The constraint on the Lorentz factor of the MeV emitting region, arising from the optical depth of the GeV photon, is rather weak.
 The weak limit does not contradict to the neutrino driven jets\citep{Aloy05} nor to the magnetic driven jets\citep{kv09,tm09}.

For a canonical set of parameters, like $z=1$,  $\delta t_M=50$s, $E_G=E_{\max}=10$GeV, $E_P=1$MeV, $L_M=10^{53}$erg, and $T_G=100$s, the constraint on the GeV region is $\Gamma_G \gtrsim 100$. 
When we apply the two zone constraint   to four  Fermi bursts, we  find   minimal  Lorentz factors of about 200-400,  about one fifth to one half (depending on $\eta$) of the  the one-zone limit  which is in the order of 1000. We conclude that one should proceed with care when applying the  one-zone limits to the Fermi data and unless we can verify  that the GeV emission is indeed produced in the same region as the lower energy prompt emission we should consider the more relaxed two zone limits.

We thank Ehud Nakar and Uri Vool for helpful discussions and an anonymous referee for helpful comments. The research was supported by an  ERC grant,  the Israel center of excellence for 
High Energy Astrophysics, a special grant of Chinese Academy of Sciences,  National basic research
programme of China grant 2009CB824800 and the National Natural Science Foundation of China under the grant  10703002 and 11073057.
TP thanks the Purple mountain observatory of  Nanjing and Huazhong University of Science and Technology for hospitality while some of this research was done.

\end{document}